\documentclass[aps,prl,amsmath,amssymb,twocolumn,superscriptaddress]{revtex4-1}  % groupedaddress
\usepackage{graphicx}  % Include figure files
\usepackage{color}
\usepackage{amssymb}   % for math
\usepackage{amsmath}
\usepackage{epstopdf}
\usepackage{natbib}
\usepackage{hyperref}  % add hypertext capabilities
\definecolor{myblue}{RGB}{46, 48,146}
\hypersetup{colorlinks=true,linkcolor=myblue,citecolor=myblue,urlcolor=blue, linktocpage}
\usepackage{bm}		   % bold math
\usepackage{braket}
\usepackage{gensymb}
\usepackage{bbold}
\usepackage{comment}
\begin{document}
\title{ Coupling of quantum-dot states via elastic-cotunneling and crossed Andreev reflection in a minimal Kitaev chain
}

%%%%%%%%%%%%%%%%%%%%%%%%%%%%%%%%%%%%%%%%%%%%%%%%%
%%%%%%%%%%%%%%%%%%%%%%%%%%%%%%%%%%%%%%%%%%%%%%%%%

\author{Zhi-Hai\! Liu }
\email{liuzh@baqis.ac.cn}
\affiliation{Beijing Academy of Quantum Information Sciences, Beijing 100193, China}

 \author{Chuanchang\! Zeng}
 \email{zengcc@baqis.ac.cn}
 \affiliation{Beijing Academy of Quantum Information Sciences, Beijing 100193, China}

 \author{H. Q. Xu}
\email{hqxu@pku.edu.cn}
 \affiliation{Beijing Key Laboratory of Quantum Devices, Key Laboratory for the Physics and Chemistry of Nanodevices, and School of Electronics, Peking University, Beijing 100871, China}
 \affiliation{Beijing Academy of Quantum Information Sciences, Beijing 100193, China}

%up to a fifth of -- individually

\begin{abstract}

Recently, exciting progress has been made in using the superconducting nanowires coupled to gate-defined quantum dots (QDs) to mimic the Kiteav chain and realize the Majorana-bound states via a poor man's route. The essential ingredient is to balance the interdot elastic-cotunneling (ECT) and crossed Andreev reflection (CAR). As theoretically proposed, this can be mediated by the Andreev bound states (ABSs) formed in the superconducting nanowires. However, most of the gate-tuning asymmetric features observed in experiments can not be captured using the current theoretical models. To address this insufficiency, here, we consider a full model that explicitly includes all the details of both the QD states and the ABSs. Remarkable agreement is found with the recent experimental observations, where our model correctly reveals the gate-tuning asymmetry in ECTs and by which the average QD state energy can also be extracted. In contrast, CARs do not depend on the tuning of QD states.   Moreover, armed with the tunability of ECTs and CARs with QD states, we also predict a controllable anisotropic superexchange interaction between  electron spins in the two separated QDs.
\end{abstract}
% and,  ,    observed in
%%%%%%%%%%%%%%%%%%%%%%%%%%%%%%%%%%%%%%%%%%%%%%%%%
%%%%%%%%%%%%%%%%%%%%%%%%%%%%%%%%%%%%%%%%%%%%%%%%%

\date{\today}
\maketitle
%%%%%%%%%%%%%%%%%%%%%%%%%%%%%%%%%%%%%%%%%%%%%%%%%
%%%%%%%%%%%%%%%%%%%%%%%%%%%%%%%%%%%%%%%%%%%%%%%%%  gate-defined
\noindent{\color{blue}\bf{Introduction}}---Semiconductor-superconductor hybrid nanostructures have recently attracted intensive attention for the exploration of nontrivial physical phenomena~\cite{Burkard2022, Flensberg2021, Frolov2020, Lutchyn2018}. Attributed to strong spin-orbit interactions (SOIs) and large Land\'{e}-$g$ factors of narrow-bandgap III-V semiconductors, superconducting nanowires have exhibited as excellent platforms for the studies of the anomalous current-phase relations~\cite{Szombati2016, Zazunov2009, Spanton2017, Laroche2019}, superconducting diode effects~\cite{Nadeem2023, Legg2022, Yokoyama2014, Su2024}, and Majorana bound states (MBSs)~\cite{Lutchyn2010, Oreg2010, Quantum2024}. When coupled with the gate-defined quantum dots (QDs), superconducting nanowires can also enable the formation of superconducting spin qubits~\cite{Hays2021, Padurariu2010, Vidal2023, Higginbotham2015, Larsen2015, Spethmann2023, Spethmann2022} and be exploited to build singlet or triplet Cooper-pair splitters~\cite{Choi2000, Deacon2015, Wang2023, Wang2022,ssy2022, Brange2021, Bordoloi2022}. More recently, it has been demonstrated that multiple QDs interconnected by a proximitized semiconductor nanowire can effectively mimic a short Kiteav chain~\cite{Sau2012,Leijnse2012,Tsintzis2022,Liu2023,Tsintzis2023,Samuelson2024,Souto2023} and thus construct MBSs in a poor-man's manner~\cite{Bordin20242,Dvir2023}. In these burgeoning implementations, the interaction between the separated QDs is found to be mediated by the Andreev bound states (ABSs) residing in the middle proximitized nanowire~\cite{Dvir2023, Bordin20242, Wang2022, Bordoloi2022, Liu2022}.

Composited as a mix of electron and hole, ABSs simultaneously facilitate the interdot elastic-cotunneling (ECT) and crossed Andreev reflection (CAR)~\cite{Bordin2024, Bordin2023}. As revealed by Refs.~\cite{Tsintzis2022, Dvir2023}, a fine-tuning reaching the sweet spot is necessarily required to balance ECT and CAR, thereby creating the poor man's MBSs. Additionally, under the interplay between SOI and the externally applied magnetic field, the breaking of spin conservation gives rise to the coexistence of spin-conserved and spin-flipped ECTs and CARs~\cite{Bordin2023}. In effect, the constructions of poor man's MBSs~\cite{Bordin20242, Dvir2023}, as well as the desired Cooper-pair splitters~\cite{Bordoloi2022, Wang2022}, demand a strict control of these spin-dependent processes. This highly involves the detailed electron states confined in  QDs and tunnel couplings with the ABSs, which, if disregarded, can not ensure a complete picture of the ABS-mediated tunneling processes. For instance, an evident gate-tuning asymmetry is observed for the spin-conserved ECT while absent for the spin-dependent CAR in the recent experiment~\cite{Bordin2023}. More
gate-tuning asymmetry appears in the magnetic field for the spin-flipped ECTs therein. These features, however, can not be well-interpreted by the theoretical models to date where the details of QD states, hence the spin-dependent tunneling processes, are not considered~\cite{Liu2022}. Therefore, it remains a top priority to clarify the effect of the confined states in QDs along with their various modulations on the ABS-mediated CARs and ECTs, as such, to facilitate the studies of rich physics in the systems.

\begin{figure}
  \centering
  \includegraphics[width=0.43\textwidth]{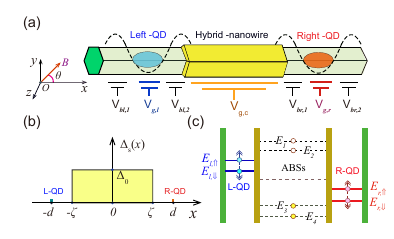}\\
  \caption{  (a) Schematic diagram of two distant gate-defined   QDs  in a semiconductor nanowire with the connection section  proximitized  by a   $s$-wave superconductor, in which $V^{}_{bl(r),1}$ and  $V^{}_{bl(r),2}$  represent the two barrier-gate potentials   for the left (right) QD, and $V^{}_{g,l(r)}$  and  $V^{}_{g,c}$  correspond to the respective plunger gate potentials. The inset depicts  the direction of  an applied magnetic field, i.e., $\mathbf{B}=B(\cos\theta,\sin\theta,0) $. (b) The  variation of the  proximity-induced  superconducting gap $\Delta^{}_{s}$   along the wire-axis ($x$) direction, in which  $2\zeta$ characterizes   the  length   of the proximitized  section  and $2d$ the  interdot  distance.  (c) The low-energy  level structures of the two-sided QDs and  middle  hybrid nanowire, with $E^{}_{l(r) \Uparrow}$ and $E^{}_{l(r) \Downarrow}$ indicating the two localized Zeeman-splitting states of the left (right) QD  and  $E^{}_{j=1-4}$   the  centered four  Andreev bound states (ABSs).  }\label{Fig1}
\end{figure}

In this Letter, we analyze the ABS-mediated tunneling processes, namely, CARs and ECTs, by constructing a tight-binding (TB) model in close analogy to the experimental setup in Ref.~\onlinecite{Bordin2023} [as schematically shown in Fig.~\ref{Fig1}(a)]. In such a TB model, all the details of the confined states are explicitly obtainable. Instead of using phenomenal parameters, we obtain the direct tunnel coupling coefficient between QDs and the intermediate ABSs strictly based on the specific localized states of each section. Utilizing the Schrieffer-Wolff transformation, we derive the analytical expressions for the amplitudes of spin-dependent ECTs and CARs that happen between the two QDs. Based on our model, we find that the ECT amplitudes are tightly dependent on the modulation of the on-site energies of the QD states, while the CAR amplitudes remain robust against these modulations. Similar dependencies in the spin-flipped ECT probabilities are also observed in the presence of a parallel magnetic field. Interestingly, our model allows us to extract the effective average on-site energy based on the observed gate-tuning asymmetry. Furthermore, we predict that, under the effect of the ECT and CAR incorporating the QD states tuning, an anisotropic superexchange interaction naturally emerges between the separated electrons confined in two QDs in the strong Coulomb-blockade regime.

%%%%%%%%%%%%%%%%%%%%%%%%%%%%%%%%%%%%%%%%%%%%%%%%%
%%%%%%%%%%%%%%%%%%%%%%%%%%%%%%%%%%%%%%%%%%%%%%%%%

%%%%%%%%%%%%%%%%%%%%%%%%%%%%%%%%%%%%%%%%%%%%%%%%%
%%%%%%%%%%%%%%%%%%%%%%%%%%%%%%%%%%%%%%%%%%%%%%%%%
%epitaxially

%  In addition,  the profiles of the local  confinement  potentials  of  are highlighted by   dashed curves.
% representing the canonical momentum operator in the axial direction we consider a semiconductor nanowire

\noindent{\color{blue}\bf{Model}}--- The two separated QDs we consider are defined around the two ends of a semiconductor nanowire with strong Rashba SOI, e.g., InAs or InSb nanowire, with the connection section proximitized by a superconductor Al,  as shown in Fig.~\ref{Fig1}(a). In the presence of a magnetic field $\mathbf{B}=B(\cos\theta,\sin\theta,0)$ applied in the $x-y$ plane,  the single-particle Hamiltonian of the hybrid nanowire along the $x$-axial direction reads as
\begin{align}
 H^{}_{e }(x) =&-\frac{\hbar^{2}_{}}{2}\frac{\partial}{\partial x}\frac{1}{ m^{}_{e}(x)}\frac{\partial }{\partial x } +\frac{g(x)\mu^{}_{\rm B}}{2} \mathbf{B}\cdot\boldsymbol{\sigma}+  V(x)\nonumber\\& -i\frac{\hbar}{2} \Big[\frac{\partial}{\partial x} \alpha(x) +\alpha(x) \frac{\partial}{\partial^{}_{x}}  \Big]\sigma^{}_{y}\ ,
   \end{align}
where $m^{}_{e}(x)$,   $g(x)$, and $\alpha(x)$ respectively correspond to the site-dependent electron effective mass  (EMS), the Land\'{e}-$g$ factor, and the Rashba  SOI strength absorbing the partial metallization effect caused by the coated superconductor. $\boldsymbol{\sigma}=(\sigma^{}_{x},\sigma^{}_{y},\sigma^{}_{z})$ denotes the Pauli matrix-vector, $\mu^{}_{\rm B}$ is the Born magneton, and $V(x)$ depicts the confinement potential induced by the gate voltages. In addition, nonzero proximity-induced superconductivity can emerge in the middle section as $\Delta^{}_{s}(x)=\Delta^{}_{0}\Theta(|x|-\zeta)$, with $\Delta^{}_{0}$ representing the induced superconducting gap, $\Theta(x)$  being a  Heaviside function, and $\zeta$  the half-width of the proximitized section [see Fig.~\ref{Fig1}(b)].

%%%%%%%%%%%%%%%%%%%%%%%%%%%%%%%%%%%%%%%%%%%%%%%%%
%%%%%%%%%%%%%%%%%%%%%%%%%%%%%%%%%%%%%%%%%%%%%%%%%

\noindent{\color{blue}\bf{Localized confined states}}---  Without loss of generality,  the local confinement potential of the left/right QD can be modeled  as $V^{}_{l/r}(x)= m^{}_{e,0}\omega^{2}_{0} (x\pm d)^{2}_{}/2+ V^{}_{g,l/r}$, in which  $m^{}_{e, 0/1}$ ($\alpha^{}_{0/1}$ and  $g^{}_{0/1}$)  quantifies the specific value of the EMS (the SOI strength  and Land\'{e} factor)  in the  bare/proximitized nanowire, $\omega^{}_{0}$       is the frequency of the harmonic potential,  $d$ is
the half interdot distance,  and $V^{}_{g,l/r}$ the relevant  plunger-gate potential.     In the case of $\Delta^{}_{z,0}\equiv g^{}_{0}\mu^{}_{\rm B} B\ll \hbar\omega^{}_{0} $,  the  lowest  Zeeman-splitting states  of the QDs  are approximated as
\begin{align}
 |\Psi^{}_{l/r,\Uparrow}  \rangle  =& \phi^{}_{ 0}  (x\pm d)  \Big[ ie^{-i\frac{x\pm d}{x_{\rm so}}_{}} \cos \frac{\vartheta}{2} |\uparrow_{ }\rangle + e^{ i \frac{x\pm d}{x^{}_{\rm so}}}\sin  \frac{\vartheta}{2}   |\downarrow^{}_{ }\rangle \Big]\nonumber\\
|\Psi^{}_{l/r,\Downarrow} \rangle= &\phi^{}_{ 0}  (x\pm d)  \Big[ ie^{-i\frac{x\pm d}{x_{\rm so}}_{}}\sin \frac{\vartheta}{2} |\uparrow_{ }\rangle- e^{ i \frac{x\pm d}{x^{}_{\rm so}}}\cos  \frac{\vartheta}{2}   |\downarrow^{}_{ }\rangle \Big]\ ,\label{PST}
 \end{align}
where $\phi^{}_{0}(x)=(\sqrt{\pi}x^{}_{0})^{-1/2}_{}
\exp[-x^{2}_{}/(2x^{2}_{0})]_{}$ is the electron spatial wave-function with $x^{}_{0}=[\hbar/(m^{}_{e,0}\omega^{}_{0})]^{1/2}_{}$ and the effective  spin-orbit length $x^{}_{\rm so  }= \hbar/(m^{}_{e,0}\alpha^{}_{0})$. Here, $\vartheta= \arccos[\sin\theta/ f^{}_{}(\theta)]$ characterizes the  spin mixing induced by SOI  with $f^{}_{}(\theta)= \! \big(\sin^{2}\theta+e^{-2x^{2}_{0}/x^{2}_{\rm so}}_{}\cos^{2}\theta \big)^{1/2}_{}$, and   $ |\uparrow^{}_{ } \rangle $ and $|\downarrow^{}_{ }\rangle$ represent the  spin-polarized states with $\sigma^{}_{ y}|\uparrow^{}_{ }\rangle =|\uparrow^{}_{ }\rangle$ and $\sigma^{}_{y}|\downarrow^{}_{ }\rangle =-|\downarrow^{}_{ }\rangle$, respectively.  Correspondingly, the  energies of the  quasi-spin states  read as  $E^{}_{ \nu, \Uparrow/\Downarrow}=   E^{}_{\nu} \pm  f^{}_{ }(\theta) \Delta^{}_{z, 0}/2 $,  with $ \nu=l,r$  and $E^{}_{\nu}=  V^{}_{g,\nu}+   (\hbar\omega^{}_{0}-m^{}_{e,0}\alpha^{2}_{0})/2$  the effective orbital  on-site energies.
 %Under the interplay between the magnetic field and the SOI,

%%%%%%%%%%%%%%%%%%%%%%%%%%%%%%%%%%%%%%%%%%%%%%%%%
%%%%%%%%%%%%%%%%%%%%%%%%%%%%%%%%%%%%%%%%%%%%%%%%%

%(relative to the  ground energy of the QDs)--the expression   of ---  facilitate the analysis of Andreev bound states (ABSs)  therein

%of  the electron and hole  single-particle Hamiltonian--$x^{ }_{1}\equiv \sqrt{\hbar/(m^{\star}_{e }\omega_{1}) }<W$-- To facilitate the analysis of   ABSs therein,
For the intermediate proximitized nanowire, the combination of the plunger-gate potential $V^{}_{g,c}$ and the two inner barrier-gate potentials of the QDs [see  $V^{}_{bl,2}$  and  $V^{}_{br,1}$ in  Fig.~\ref{Fig1}(a)], in principle, can also endow a concave-shaped confinement, $V^{}_{in}(x)=  m^{}_{e,1}\omega^{2}_{1}x^{2}_{}/2+ V^{}_{g,c}$  with   $\omega^{}_{1}$  specifying the parabolic potential. %, in which  $\omega^{}_{1}$  specifies  the  parabolic potential    with   $x^{}_{1}=[\hbar/(m^{}_{e,1}\omega^{}_{1})]^{1/2}_{}\!<\! \zeta$. %.
Then, the effective  Bogoliubov-de Gennes (BdG) Hamiltonian for the proximities section   can be expressed as
$  H   ^{}_{\rm BdG } = [p^{2}_{x}/(2 m ^{  }_{e,1})+\alpha^{  }_{ 1 }p^{}_{x}\sigma^{}_{y}+ V^{}_{in}(x)]\tau^{}_{z} +\Delta^{}_{0}\tau^{}_{x} +  \Delta^{}_{z,1}(\cos\theta\sigma^{}_{x} +\sin\theta\sigma^{}_{y})/2$,  with  $p^{}_{x}=-i\hbar\partial/(\partial x)$,  $\Delta^{}_{z,1}=g^{ }_{1} \mu^{}_{\rm B} B$, and $\tau^{}_{x,y,z}$ the  Pauli matrices defining in the particle-hole (PH) space. As such, the low-energy ABSs of the intermediate nanowire can be analytically derived based on the effective Hamiltonian~\cite{Supplementary}. The energies of  the lowest two ABSs above the Fermi (zero-energy) level take the form of
 %obtained as
 \begin{align}
E^{}_{1/2}=  \sqrt{\mu^{2}_{}+\Delta^{2}_{0} }\pm \frac{ f^{ }_{1}(\theta)}{2}  \Delta^{}_{z,c}\ ,
\label{ENS}
\end{align}
% acting as the effective chemical potential of the hybrid nanowire
 with $ \mu^{}_{}= V^{}_{g,c}+(\hbar\omega^{}_{1} -m^{ }_{e,1}\alpha^{2}_{1} )/2$  and $f^{ }_{ 1}(\theta)$  similar to  $f^{ }_{ }(\theta)$ but with  $x^{}_{0}$ and $x^{}_{\rm so}$
replaced by $x^{}_{1}=[\hbar/(m^{}_{e,1}\omega^{}_{1})]^{1/2}_{}$ and $x^{\prime}_{\rm so}=\hbar/(m^{}_{e,1}\alpha^{}_{1})$, respectively.  The energies of the nearest two ABSs below the Fermi level [as seen in Fig.~\ref{Fig1}(c)]  are given by $E^{}_{3/4}=-E^{}_{2/1}$,  in accordance with the PH symmetry of the BdG Hamiltonian $ {\cal P}^{}_{} H ^{}_{\rm BdG}{\cal P}^{-1}_{}= -H ^{}_{\rm BdG}$,  with ${\cal P}^{}_{}=i\tau^{}_{y}\sigma^{}_{y}   K $     and $  K $ being the PH and complex-conjugation operators, respectively. Due to the coexistence of the SOI and the proximity effect,  they represent quasi-particles as a combination of electron and hole components characterized by different spin mixings. Hereafter, they are denoted by the  Bogoliubov quasi-particle operators  $\gamma^{}_{j}$ with $j=1-4$ respectively.

The extension of the  ABSs in the bare section leads to direct tunnel-couplings with the  QD states as defined in Eq.~(\ref{PST}). Let's denote the creation (annihilation) operator for the quasi-spin state $|\Psi^{}_{l/r, s}\rangle$ with $s=\Uparrow,\Downarrow$ as $d^{\dagger}_{l/r s}$ ($d^{}_{l/r s}$), then the corresponding  tight-binding (TB) Hamiltonian for the hybrid  nanostructure can be  formulated as
 %\begin{small}
 \begin{align}
H^{}_{\rm TB}= &\sum^{}_{\nu, s} E^{}_{\nu, s} d^{\dagger}_{\nu s} d^{}_{\nu s}
 +\sum^{ }_{j } E^{}_{j}\gamma^{\dagger}_{j}\gamma^{}_{j}\nonumber\\ &
~ +\sum^{}_{ \nu,s,j }    \left( t^{}_{\nu s,j} d^{\dagger}_{\nu s} \gamma^{}_{j}  +{\rm h.c.} \right)\ ,\label{tun}
 \end{align}
 %\end{small}
in which $ t ^{}_{\nu s,j}$  characterizes the tunnel-coupling amplitude from the QD state $|\Psi^{}_{\nu, s}\rangle$ to the  ABS  $\gamma^{}_{j}$. Specifically, in addition to  a structure-dependent (spinless) parameter $t^{}_{0}$, because of SOI, $t^{}_{\nu s, j}$ also depends on the degrees of spin mixing in the two localized states and the accumulated spin rotation phase $\Phi^{}_{\rm so}= \tilde{d}/x^{}_{\rm so}$ in the tunneling.  Note that, the effective length $\tilde{d}$ is different from the interdot distance $d$,  due to the  SOI inhomogeneity along the axial direction. The details of the direct tunneling amplitude  $ t ^{}_{\nu s,j}$  are presented in the Supplementary~\cite{Supplementary}.

% , and
%reformulated
%in this subset
%it is found that = \sum^{}_{j=1,2} (t^{\prime  }_{ l\sigma^{\prime },j } t^{\prime \ast}_{r\sigma,j}- t^{}_{l\sigma^{\prime },j } t^{\ast}_{r\sigma,j})/E^{}_{j}~~~
%In fact, the on-site energies of the    states in the  left and right QDs  can be  individually   tuned by regulating the plunger-gate potentials, $V^{}_{g,l}$ and $V^{}_{g,r}$.

  % Evidently,  the    energy of the ABSs  $E^{}_{j}$    can be regulated by the middle plunger-gate potential   $V^{}_{g, c}$ based on Eq.~(\ref{ENS}). -electrons in the

\begin{figure}
  \centering
  \includegraphics[width=0.45\textwidth]{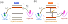}\\
  \caption{ (a)  Schematic diagram of a  ECT, in which there is an electron   quasi-spin  state transferring  from the right-QD to the left-QD as mediated by the ABS  $\gamma^{}_{2}$ of the middle hybrid nanowire. (b)  Schematic  of  a   CAR with  two  electron   quasi-spin  states creating  in the two-sided  QDs, and under  the intermediation of    two middle  PH-symmetric   ABSs, i.e., $\gamma^{}_{2}$ and $\gamma^{}_{3}$.   Here, $\mathbf{t}^{}_{l(r) ,j}$  indicate the direct tunnel-coupling amplitude vector between the left (right) QD states and the ABS $\gamma^{}_{j}$, and is  comprised by $\mathbf{t}^{}_{l(r) ,j}=\{t^{}_{l(r)\Uparrow,j},t^{}_{l(r)\Downarrow,j}\}^{ }_{}$ .          }
  \label{Fig2}
\end{figure}

\noindent{\color{blue}\bf{Effective interdot interactions}}---  When the QD states are largely detuned from the ABSs, i.e., $ E^{}_{\nu,s}\ll |E^{}_{j}|$,   it is interesting to find that the direct tunnel-couplings in Eq.~(\ref{tun}) can simultaneously facilitate the interdot elastic-cotunneling (ECT)  and crossed Andreev reflection(CAR). Practically, this is accompanied by virtual excitations or evacuations of the middle higher- or lower-energy ABSs,  as seen in Fig.~\ref{Fig2}. Through using the Schrieffer-Wolff transformation to eliminate the intermediate processes, the  effective  Hamiltonian for the two long-range interactions can be found as  $H^{}_{\rm ECT}=\sum{}_{ss^{\prime}_{}=\Uparrow,\Downarrow} \big(d^{\dagger}_{ls}{\rm T}^{}_{ss^{\prime}_{}}d^{}_{r s^{\prime}_{}}+{\rm h.c.}\big)$ and   $H^{}_{\rm CAR}=\sum{}_{ss^{\prime}_{}} \big(d^{\dagger}_{ls}{\rm R}^{}_{ss^{\prime}_{}}d^{\dagger}_{r s^{\prime}_{}}+{\rm h.c.}\big)$, with the  spin-dependent amplitudes evaluated by
 %\begin{small}
 \begin{align}
 {\rm T}^{}_{ss^{\prime}_{}}= &\sum^{}_{j=1-4}\frac{t^{}_{ls,j}t^{\ast}_{rs^{\prime}_{},j}}{2}
 \left(\frac{1}{E^{}_{l,s  }-E^{}_{j} }+\frac{1}{E^{}_{r,s^{\prime}_{} }-E^{}_{j}} \right) \nonumber\\
 {\rm R}^{}_{ss^{\prime}_{}}=& \sum^{}_{j=1-4} \frac{t^{}_{ls,j}t^{}_{rs^{\prime}_{},  (5-j)}}{2} \left(\frac{1}{E^{}_{l,s}-E^{}_{j}}+\frac{1}{E^{}_{ (5-j) }-E^{}_{r,s^{\prime}_{}}}\right)\ .\label{TRS}
 \end{align}
Intuitively, as can be seen above, ECT is achieved via the sequential single-electron ($\gamma^{}_{j}$) transports. In contrast, CAR involves the creation or annihilation of two electrons in the QDs, mediated by the PH-symmetric ABSs $\gamma^{}_{j}$  and $\gamma^{}_{5-j}$ originating from a Cooper pair [see Fig.~\ref{Fig2}(b)]. We want to stress that, the explicit expressions of  $t^{}_{\nu s, j}$ and $E^{}_{\nu,s}$ are all obtainable in our model, hence allowing the study of the CARs and ECTs in a complete sense. In what follows, we discuss this in more detail.

\begin{figure}
  \centering
  \includegraphics[width=0.48\textwidth]{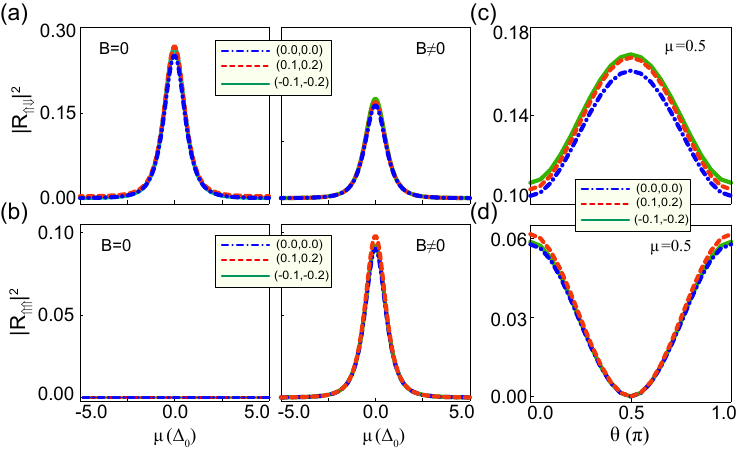}\\
  \caption{ (a)  The   absolute values of the spin-flipped CAR amplitude   $|{\rm R}^{}_{\Uparrow\Downarrow}|^{2}_{}$,   in  units of $4t^{4}_{0}/\Delta^{2}_{0}$,  as a function of $\mu^{}_{ }$ with   different  on-site energies $E^{}_{l}$ and $E^{}_{r}$ (see the inset), and $\theta=0$.  The    right (left) panel corresponds to the case  with  a zero (nonzero) magnetic field.  (b) Same as (a), but show the   trends of the spin-conserved quantity $|{\rm R}^{}_{\Uparrow\Uparrow}|^{2}_{}$ in different conditions.  (c) and (d)  At a fixed  $\mu$,  the variations    $|{\rm R}^{}_{\Uparrow\Downarrow}|^{2}_{}$ and $|{\rm R}^{ }_{\Uparrow\Uparrow}|^{2}_{}$ versus $\theta$ in the presence of the  magnetic field and with different on-site energies.   Here,  the energies  $E^{}_{l } $, $E^{}_{r}$, and the chemical potential $\mu$  are in units of $\Delta^{}_{0}$,     the nonzero magnetic field is characterized by $\Delta^{}_{z,0}=0.1\Delta^{}_{0}$, and   the other parameters are fixed as   $\tilde{d}^{}_{ }=200$~nm,  $x^{}_{\rm so}= 160$~nm, $x^{\prime}_{\rm so}=240$~nm, $x^{}_{0}=30$~nm, $x^{}_{1}=70$~nm, and  $\Delta^{}_{z,c}=0.5\Delta^{}_{z,0}$ in the simulations. }\label{Fig3}
\end{figure}

%\textcolor{red}{Due to the coordination of the paired ABSs in forming CAR,} it is evident that     $ {\rm R}^{}_{ss^{\prime}_{}}$ in Eq.~(\ref{TRS}) is mainly dependent on the effective chemical potential of the ABSs $\mu$, and less susceptible to the variation of the QD states or the applied magnetic-field strength. In this scenario,
By noting the energy conservation warranted by the paired ABSs in $ {\rm R}^{}_{ss^{\prime}_{}}$, i.e., $E_j + E_{5-j} \equiv 0$, the Hamiltonian for the CAR interaction can be effectively formulated as
% using Under this circumstance if there are some specific conditions given
\begin{align}
H^{}_{\rm CAR}=   -\frac{t^{2}_{0}\Delta^{}_{0}}{\mu^{2}_{}+\Delta^{2}_{0}} \left(\begin{array}{c}
  d^{}_{l\Downarrow}\\
  -d^{}_{l\Uparrow}
  \end{array}\right)^{\dagger}_{}\hat{\mathbf{U}}^{}_{\rm   } ( 2\Phi^{}_{\rm so} ) \left(\begin{array}{c}
  d^{\dagger}_{r \Uparrow}\\
  d^{\dagger}_{r \Downarrow}
  \end{array}\right)^{ }_{}+{\rm h.c.}\ ,
  \label{CARS}
 \end{align}
with $\hat{\mathbf{U}}^{}_{}(2\Phi^{}_{\rm so})= \exp[-2i \Phi^{}_{\rm so } \hat{ n }\cdot \boldsymbol{\sigma}]$ representing the spin rotation matrix induced by SOI  and $\hat{ n }=(\sin  \vartheta, 0,\cos \vartheta )^{}_{}$ the direction vector the rotation axis. In this form, the $\mu^{}_{ }$-dependence of the amplitude ${\rm R}^{}_{ss^{\prime}_{}}$ is thus directly revealed by the coefficient of Eq.~(\ref{CARS}) now. Despite of the values of $E^{}_{\nu }$ and $B$, as shown in Fig.~\ref{Fig3}(a), $|{\rm R}_{ s \bar{s}^{ }_{}}|^{2}_{}$   manifests a rather robust dependency of $\mu^{}_{ }$, i.e., decreasing with increasing $|\mu|$, and always attain its maximum at $\mu^{}_{ }=0$. On the other hand, Fig.~\ref{Fig3}(c) shows the dependence on the specific expression of $\hat{n}$, where robustness against the on-site energies of the QD states ($E^{}_{l,r }$) can also be observed. Similar results for $|{\rm R}^{}_{ss^{ }_{}}|^{2}_{}$ are plotted in Figure~\ref{Fig3}(b, d), except at $B=0$ or  $\theta=\pm \pi/2$ where the spin-conserved CAR is suppressed because of the spin conservation. Note that the above equation is explicitly independent of the on-site energy of the QD states; therefore, for CARs, the main features are reserved even without involving the effects of the QD states. This explains why the model utilized in Ref.~\cite{Liu2022} is sufficient to interpret the observed CARs in experiments, which, however, is definitely not for the ECTs, as shown following.

%%shows that the spin-conserved  quantity $|{\rm R}^{2}_{ s \bar{s}^{ }_{}}|$ exhibits a  similar tendency, except when $B=0$ or  $\theta=\pm \pi/2$.   The suppression of the spin-conserved CAR   in this case is because of  the  spin conservation, i.e.,  $[H^{}_{e }(x),\sigma^{}_{y}]=0$.  With the change of the  magnetic-field direction,
%Fig.~\ref{Fig3}(c) and \ref{Fig3}(d) show  that  $|{\rm R}^{2}_{s\bar{s}^{}_{}}|$  and $|{\rm R}^{2}_{ss^{ }_{}}|$  will be varied   on the basis of the specific expression of $\hat{n}$, and the specific values are almost insensitive to the change of the on-site energies.
% All of these are consistent with the   experimental results in Refs.~\onlinecite{Liu2022,Bordin2023}.

%In contrast,
%because the ECT is contributed by single-electron transports,
%the spin-dependent amplitude ${\rm T}^{}_{ss^{\prime}_{}}$ in Eq.~(\ref{TRS}) can be effectively regulated by the on-site energies. %Based on the explicit forms of  $t^{}_{\nu s, j}$ and $E^{}_{\nu,s}$,   it
%After plunging in the explicit forms of  $t^{}_{\nu s, j}$ and $E^{}_{\nu,s}$,
Similarly, ${\rm T}^{}_{ss^{\prime}_{}}$ for ECT is further reformulated as,
  \begin{align}
  {\rm T}^{}_{ss^{\prime}_{}}= &- \frac{t^{2}_{0}( \mu^{}_{}+E^{}_{ss^{\prime}_{}} )}{ \mu^{2}_{}+\Delta^{2}_{0} }  \left[\hat{\mathbf{U}}^{}_{\rm   } (-2\Phi^{}_{\rm so} ) \right]^{}_{ss^{\prime}_{}} +\frac{t^{2}_{0}\Delta^{}_{z,c}f^{ }_{1}(\theta) }{2(\mu^{2}_{}+\Delta^{2}_{0})} \nonumber\\
 &\times  \left\{  [ \mathbf{ \Pi}^{}_{\rm  }(\theta)\cdot \boldsymbol{\sigma} ]^{}_{ss^{\prime}_{}} +i\delta^{}_{ss^{\prime}_{}}\sin(2\Phi^{}_{\rm so})\cos\varphi \right\}\ ,
 \label{TSS}
  \end{align}
with $E ^{}_{ ss^{\prime}_{}}=(E^{}_{l,s}+E^{}_{r, s^{\prime}_{}})/2$, $\delta^{}_{ss^{\prime}_{}}$ being the Kronecker delta function, $\varphi=\arccos[\sin\theta/ f^{}_{1}(\theta)]$, and $\mathbf{\Pi}^{}_{\rm }(\theta)$ is a parameter vector~\cite{Supplementary}. %Here, $\mathbf{\Pi}^{}_{\rm }(\theta)=(\Pi^{}_{x},0,\Pi^{}_{z}) $ is  a parameter vector, with the nonzero components   $\Pi^{}_{x}=\cos^{2}_{}\Phi^{}_{\rm so} \sin(\vartheta- \varphi)-\sin^{2}\Phi^{}_{\rm so} \sin (\vartheta+\varphi)$ and  $\Pi^{}_{z}= \cos^{2}_{}\Phi^{}_{\rm so} \cos(\vartheta- \varphi)-\sin^{2}\Phi^{}_{\rm so} \cos (\vartheta+\varphi)$.

\begin{figure}
  \centering
  \includegraphics[width=0.48\textwidth]{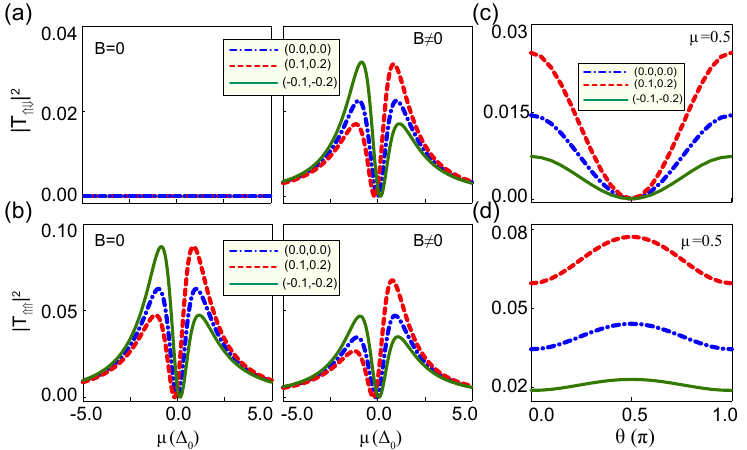}\\
  \caption{ (a)-(b)      The absolute values of the spin-dependent ECT amplitudes between the double QDs  in Fig.~\ref{Fig3}, $|{\rm T}^{ }_{\Uparrow\Downarrow}|^{2}_{}$   and $|{\rm T}^{}_{\Uparrow\Uparrow}|^{2}_{}$, in units of $4t^{2}_{0}/\Delta^{2}_{0}$, as a function of $\mu^{}_{ }$, with different on-site energies  ($E^{}_{l}$ and $E_{r}$) and  magnetic-field strength ($B$), and the direction angle $\theta=0$.  (c)-(d) The variations of $|{\rm T}^{}_{\Uparrow\Downarrow}|^{2}_{}$ and $|{\rm T}^{}_{\Uparrow\Uparrow}|^{2}_{}$ versus $\theta$, in the case of different on-site energies and  with $\Delta^{}_{z,0}=0.1\Delta^{}_{0}$  and   $\mu=0.5\Delta^{}_{0}$.  }\label{Fig4}
\end{figure}

Admitted by the spin conservation with $B=0$, the spin-flipped ECTs are forbidden and hence we have ${\rm T}^{}_{s\bar{s}}=0$, as shown in Fig.~\ref{Fig4}(a). Contrarily, the spin-conserved quantity $|{\rm T}^{}_{ss}|^{2}_{}$ exhibit evident $\mu$- and $E^{}_{l,r}$-dependencies, as can be seen from Fig.~\ref{Fig4}(b). It is found $|{\rm T}^{}_{ss}|^{2}_{}$ vanishes at $\mu\simeq -E^{}_{ss}$, accompanied with two local peaks emerging around. Interestingly, based on Eq.~(\ref{TSS}), the ratio between these two peak values can be evaluated as $\gamma^{}_{0}=  [E^{}_{ss}+ (E^{2}_{ss}+\Delta^{2}_{0})^{1/2}_{}]^{2}_{}/ [E^{}_{ss}- (E^{2}_{ss}+\Delta^{2}_{0})^{1/2}_{}]^{2}_{}$. As a result, the spin-conserved tunneling probability is generically $\mu$-asymmetric. This, remarkably, can well capture the experimental observations in Ref.~\onlinecite{Bordin2023} [see Fig.~2(f) therein], according to which estimations of $\gamma_{0}\simeq  2.4$ and the average on-site energies $E^{}_{ss}\simeq0.22 \Delta^{}_{0}$ can thus be extracted. We want to emphasize that, the above significant prediction we found for the spin-conserved ECT has not been discussed anywhere yet,  which we believe is ascribed to the inclusion of the nonzero on-site energies considered in our model.

Note that, the asymmetric $\mu$-dependence of $|{\rm T}^{}_{ss}|^{2}_{}$ can also appear when a parallel magnetic field ($B\neq 0$) is applied, even when $E^{}_{ss}=0$, see the blue dashed line in Fig.~\ref{Fig4}(b). Similarly, spin-flipped ECT comes to play in the presence of $B\neq 0$ as well, presenting a much similar $\mu$-dependence as that for $|{\rm T}^{}_{ss}|^{2}_{}$ at $B =0$ [right panel, Fig.~\ref{Fig4}(a)]. This again, when considering $E^{}_{s\bar{s}}\neq 0$, can well explain what has been observed for ${\rm T}^{}_{s\bar{s}}$ in the experiment, see Fig.~3(d) of Ref.~[\onlinecite{Bordin2023}]. Moreover, the spin-dependent ECT amplitudes also vary with the direction of the magnetic field. As shown in  Figs.~\ref{Fig4}(c, d), for the case of $ \mu^{}_{}> E_{\nu, s} $, $|{\rm T}^{}_{s\bar{s}}|^{2}_{}$ and  $|{\rm T}^{}_{s s }|^{2}_{}$ attain their local maximum and minimum values at $\theta=0,\pi$, respectively, showing a significant modulation by the on-site energies. %More features for the $\mu$-asymetry dependence of the spin-conserved ECTs in magnetic fields can be well captured in our scheme, which is, however, obviously beyond the scope of the previously studied models~\cite{Supplementary}.
%but with the specific values highly sensitive to the on-site energies of the QD states involved in the processes.

\begin{figure}
\includegraphics[width=0.46\textwidth]{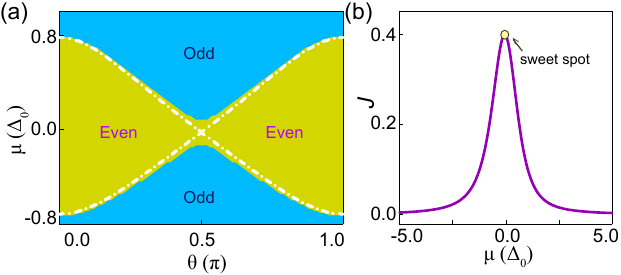}
\caption{(color online) (a) The   parity-phase diagram of the ground state of the double QDs   in the  $\theta- \mu^{}_{}$ plane, with $E^{}_{l,r \Downarrow}=0$,   $\Delta^{}_{z,0}=0.2$, $U=5.0$  (in a units of $\Delta^{}_{0}$) and   the other system parameters  as same as that in  Fig.~\ref{Fig3}.  In addition,  the dot-dashed  (white) curves indicate  the correlation of the two modulation parameters  to attain $|{\rm T}^{}_{\Downarrow\Downarrow}|^{2}_{}=|{\rm R}^{}_{\Downarrow\Downarrow}|^{2}_{}$. (b) Within the strong Coulomb-blockade regime of $E^{}_{l}=-0.3 $,  $E^{}_{r}=-0.4$,  and $\Delta^{}_{z,0}=0.05$, the superexchange interaction  strength $J$  in Eq.~(\ref{Jstr}), in a unit of $4t^{4}_{0}/\Delta^{2}_{0}$,   as a function of  $\mu^{}_{ }$. }
\label{Fig5}
\end{figure}

\noindent{\color{blue}\bf{More physics relates to on-site energy}}--- As has been addressed in Ref.~[\onlinecite{Tsintzis2022}], the formation of poor man's MBSs demands double QDs with symmetric on-site energy and with the low-energy Zeeman splitting states kept around the Fermi level, i.e., $E^{}_{\nu,\Downarrow}\simeq 0$. Being a zero-energy and half-fermion state, the emergence of a pair of the MBSs can also be manifested by the parity degeneracy in the ground state of the double QDs. Figure~\ref{Fig5}(a) shows the distribution of the parity of the ground state in the $ \theta-\mu^{}_{ }$ plane, where, for completeness, the effect of intradot Coulomb repulsions is also included in the calculation~\cite{Supplementary}. Clearly, the boundary between the odd-parity and even-parity areas indicates the correlation of the two modulation parameters required to attain the parity degeneracy. This is consistent with the conventional criteria for the poor man's MBSs in Ref.~[\onlinecite{Leijnse2012}], where a balance between the interdot spin-conserved ECT and CAR $|{\rm T}^{}_{\Downarrow\Downarrow}|^{2}_{}=|{\rm R}^{}_{\Downarrow\Downarrow}|^{2}_{}$ is necessary. Note that, as there is no spin-conserved CAR at $ \theta=\pi/2$ (see Fig.~\ref{Fig3}(d)), the emergence of parity degeneracy does not really ensure the creation of the MBSs, as indicated by the deviation emerging there in Fig.~~\ref{Fig5}(a).

When tuned into the strong Coulomb blockade regime, i.e., $|E^{}_{\nu,s}|,|E^{}_{j}|\ll U$ with $U$ denoting the intradot Coulomb repulsion strength, the electron occupancy of the double QDs is fixed as (1,1), i.e., each dot is now occupied by a single electron. In this scenario, the electrons of the two QDs will establish an effective exchange interaction via the interdot CAR and ECT, as given in Eqs.~(\ref{CARS}) and (\ref{TSS}). Explicitly, at the limit of $|E^{}_{\nu }/\Delta^{}_{z,s}|\gg 1$, the superexchange Hamiltonian is found as,
  ${\cal H}^{}_{\rm eff}=   J \Big[ \cos^{2}_{} (2\Phi^{}_{\rm so}) \mathbf{S}^{}_{l}\cdot \mathbf{S}^{}_{r}  +\sin (4\Phi^{}_{\rm so}) \hat{ n} \cdot (\mathbf{S}^{}_{l}\times \mathbf{S}^{}_{r})+\sin^{2}_{}(2\Phi_{\rm so}) \mathbf{S}^{}_{l} \overleftrightarrow{\Gamma} \mathbf{S}^{}_{r} \Big]$~\cite{Supplementary}. Here, we have $\overleftrightarrow{\Gamma}=2\hat{ n }\hat{ n}-1$, $\mathbf{S}^{}_{ \nu=l,r}=  ( 1/2) \sum^{}_{s,s^{\prime}_{}}d^{\dagger}_{\nu s^{\prime}_{}}\boldsymbol{\sigma}^{}_{s^{\prime}_{}s^{}_{}}
  d^{}_{\nu s} $ representing the electron quasi-spin operators, and  the strength  of the exchange interaction    %\begin{align}
 \begin{align}
    J = \frac{ t^{4}_{0}U}{(\mu^{2}_{ }+\Delta^{2}_{0})^{2}_{}} \Big[\frac{ (2\mu^{ }_{ }+E^{}_{+})^{2}_{} }{U^{2}_{}- E^{2}_{-}}-\frac{4\Delta^{2}_{0}}{E^{}_{+}(2U+ E^{}_{+})}\Big]\ ,
   \label{Jstr}
    \end{align}
with  $ E^{}_{\pm}=E^{}_{l}\pm E^{}_{r} $.  It is evident that the direction of the Dzyaloshinskii-Moriya vector in ${\cal H}^{}_{\rm eff}$, viz., $\hat{  n }$, vary with the direction of the magnetic field. Moreover, the exchange strength $J $ can be regulated by the chemical potential $\mu^{}_{ }$, as shown in Fig.~\ref{Fig5}(b). When $\mu^{}_{ }$ is tuned approaching to (away from) the zero point, $J$ is effectively switched on (off). Specifically,  at $\mu=0$, it is found that the maximum of $  J $ is achieved and the modulation of the exchange interaction reaches a sweet point as $\partial   J  / \partial  \mu^{}_{}=0$. We believe this is important in studying the long-range spin transfer and exchange gate using the QD-coupled superconducting nanostructures.

%change of the chemical potential. %in the modulation.
%  the direction of the     Dzyaloshinskii-Moriya  vector  $\hat{\boldsymbol{n}}$ can be effectively regulated by the magnetic field direction, and

\noindent{\color{blue}\bf{Conclusion---}} Focusing on the interaction between two separated QDs mediated by the ABSs of a superconducting proximitized nanowire,  we clarify different interesting effects on the interdot ECTs and CARs from tuning the on-site energies of the QD states, as explicitly considered in our model. Regarding the CARs, we demonstrate that the applied modulations do not affect the spin-dependent tunneling amplitudes. This validates the sufficiency of the simplified treatment in Ref.~[\onlinecite{Liu2022}], in which the specific on-site energies of the QD states are not accounted for. However, the lack of interpretation of the $\mu$-assymetry for ECTs, as observed in the experiment~\cite{Bordin2023}, immediately demands a theoretical model going beyond that used in Ref.~[\onlinecite{Liu2022}].

Instead, the complete TB model considered in this work can correctly capture all the experimental observations~\cite{Bordin2023}. Besides the predictions of those $\mu$-assymetry for ECTs with or without the magnetic field (as mentioned above), within our scheme, average on-site energy can be effectively extracted according to the experimental observations~[\onlinecite{Bordin2023}]. We further demonstrate that our model allows for exploring some other intriguing phenomena, such as the poor man's MBSs via a parity phase diagram and a controllable anisotropic superexchange interaction between electrons. Therefore, we believe our work will facilitate more interesting studies of the rich physics in this system.
% We demonstrate that  the robustness of the

%%%%%%%%%%%%%%%%%%%%%%%%%%%%%%%%%%%%%%%%%%%%%%%%%
%%%%%%%%%%%%%%%%%%%%%%%%%%%%%%%%%%%%%%%%%%%%%%%%%

% in the regulation of the exchange interaction strength.

%%%%%%%%%%%%%%%%%%%%%%%%%%%%%%%%%%%%%%%%%%%%%%%%%
%%%%%%%%%%%%%%%%%%%%%%%%%%%%%%%%%%%%%%%%%%%%%%%%%

%%%%%%%%%%%%%%%%%%%%%%%%%%%%%%%%%%%%%%%%%%%%%%%%%
%%%%%%%%%%%%%%%%%%%%%%%%%%%%%%%%%%%%%%%%%%%%%%%%%

%%%%%%%%%%%%%%%%%%%%%%%%%%%%%%%%%%%%%%%%%%%%%%%%%
%%%%%%%%%%%%%%%%%%%%%%%%%%%%%%%%%%%%%%%%%%%%%%%%%

%%%%%%%%%%%%%%%%%%%%%%%%%%%%%%%%%%%%%%%%%%%%%%%%%
%%%%%%%%%%%%%%%%%%%%%%%%%%%%%%%%%%%%%%%%%%%%%%%%%

%%%%%%%%%%%%%%%%%%%%%%%%%%%%%%%%%%%%%%%%%%%%%%%%%
%%%%%%%%%%%%%%%%%%%%%%%%%%%%%%%%%%%%%%%%%%%%%%%%%

%%%%%%%%%%%%%%%%%%%%%%%%%%%%%%%%%%%%%%%%%%%%%%%%%
%%%%%%%%%%%%%%%%%%%%%%%%%%%%%%%%%%%%%%%%%%%%%%%%%

  % Here, we demonstrate the  correspondence is  not exclusive to the topologically nontrivial classes, but also can be extended to  the trivial  symmetry classes.

 %%%%%%%%%%%%%%%%%%%%%%%%%%%%%%%%%%%%%%%%%%%%%%%%%
%%%%%%%%%%%%%%%%%%%%%%%%%%%%%%%%%%%%%%%%%%%%%%%%%

\textit{Acknowledgment.---} We thank for useful discussion with D. Loss,  L.  Kouwenhoven, Y. Zhou, J-Y Wang, and P. Zhang. This work is supported by the National
Natural Science Foundation of China (Grants No. 92165208
and No. 11874071). C. Z. acknowledges support from NSFC (Grant No. 12104043).

%%%%%%%%%%%%%%%%%%%%%%%%%%%%%%%%%%%%%%%%%%%%%%%%%%%%%%%%%%%
%%%%%%%%%%%%%%%%%%%%%%%%%%%%%%%%%%%%%%%%%%%%%%%%%%%%%%%%%%%

%%%%%%%%%%%%%%%%%%%%%%%%%%%%%%%%%%%%%%%%%%%%%%%%%%%%%%%%%%%%%

%%%%%%%%%%%%%%%%%%%%%%%%%%%%%%%%%%%%%%%%%%%%%%%%%
%%%%%%%%%%%%%%%%%%%%%%%%%%%%%%%%%%%%%%%%%%%%%%%%%

\end{document}